# Evidence of metallic clustering in annealed $Ga_{1-x}Mn_xAs$ from atypical scaling behavior of the anomalous Hall coefficient

H.K. Choi, W.O. Lee, Y.S. Oh, K.H. Kim, and Y.D. Park[a]
*CSCMR & School of Physics, Seoul National University, Seoul 151-747 Korea*

S.S.A. Seo and T.W. Noh
*ReCOE and School of Physics, Seoul National University NS50, Seoul 151-747 Korea*

Y.S. Kim, Z.G. Khim, I.T. Jeong, and J.C. Woo
*School of Physics, Seoul National University NS50, Seoul 151-747 Korea*

S.H. Chun
*Department of Physics and Institute of Fundamental Physics, Sejong University, Seoul 143-747 Korea*

We report on the anomalous Hall coefficient ($R_s$) and longitudinal resistivity ($\rho_{xx}$) scaling relationship ($R_S \propto \rho_{xx}^n$) on a series of annealed $Ga_{1-x}Mn_xAs$ epilayers ($x \approx 0.055$). As-grown samples exhibit scaling parameter $n$ of ~ 1. Near the optimal annealing temperature, we find $n \approx 2$ to be consistent with recent theories on the intrinsic origins of anomalous Hall Effect in $Ga_{1-x}Mn_xAs$. For annealing temperatures far above the optimum, we note $n > 3$, similar behavior to certain inhomogeneous systems. This observation of atypical behavior agrees well with characteristic features attributable to spherical resonance from metallic inclusions from optical spectroscopy measurements.

Ever since the first reports of carrier mediated ferromagnetic ordering in III-V diluted magnetic semiconductors (DMS), anomalous Hall Effect (AHE) measurements have had an important role in their characterization[1]. First, observation of AHE was tacitly believed to be attributed from a single phase carrier mediated DMS materials[2,3]. AHE has been utilized to indirectly measure magnetic properties, especially where direct magnetization measurements are difficult, with novel demonstrations of carrier mediated ferromagnetic ordering manipulated by electric fields[4] and even with reports of AHE near room temperature for GaAs-based DMS systems[5]. Whether observations of AHE are indeed unique to a single phase DMS materials is best illustrated in AHE measurements of $Ti_{1-x}Co_xO_2$ system[6]; whether to carrier-mediated DMS materials has been studied in both Mn:Ge system[7] as well as digitally doped Mn/GaAs system[8].

The origins of DMS magnetic ordering, especially whether the carriers are spin-polarized or the magnetization is due to secondary phases, are essential issues to be considered for the applicability of DMS for spintronic device applications. Even for $Ga_{1-x}Mn_xAs$, the seminal III-V DMS, there have been reports of secondary phases which may contribute to the observed magnetic properties[9] as well as localization of spin-polarized carriers near the magnetic impurity[10]. Recently, a means to increases magnetic coercivities of $Ga_{1-x}Mn_xAs$ by inclusion of small concentration of nanometer-sized MnAs has been reported[11]. AHE in $Ga_{1-x}Mn_xAs$ has been studied theoretically by many groups with its origins ranging from Berry phase in the momentum space[12] to phonon-assisted hopping of holes between localized states in the impurity band[8,13]. In this letter, we illustrate the sensitivity of the Hall Effect measurements to metallic inclusions within the $Ga_{1-x}Mn_xAs$ host by atypical scaling relationship of the anomalous Hall coefficient ($R_S$) to resistivity ($\rho_{xx}$).

Several 100 nm thick $Ga_{1-x}Mn_xAs$ samples ($x \approx 0.055$) are prepared by LT-MBE on epi-ready SI GaAs(001) substrates after 500 nm GaAs buffer layers are first grown, which details are reported elsewhere[14]. After growth, high-resolution x-ray diffraction (HRXRD) measurements is used to verify $x$ of 0.052, 0.052, and 0.056 for sample A, sample B, and sample C, respectively. The samples are fashioned into electrically isolated 300 μm x 1900 μm Hall bar structures. As-grown, the magnetic ordering temperature ($T_C$), as estimated by SQUID magnetometry and/or by transport measurements, is found to be 50 - 62 K. Samples are then annealed in a tube furnace in a flowing dry $N_2$ environment for one hour with annealing temperature ($T_A$), measured by a thermocouple near the sample, ranging from 200°-400° C. After annealing, indium contacts are fashioned and verified as ohmic for transport measurements in a closed-cycle cryostat and/or in a Quantum Design PPMS with customized ac lock-in technique capabilities.

Our observed effects due to annealing on the transport and magnetic properties are similar to those reported by others[15-17]. Resistivities of the three samples initially decrease (along with corresponding increase in $T_C$) with increasing $T_A$ as donor impurities such as Mn interstitial concentrations (Mn$_I$) are reduced by out-diffusion to the surface and passivated[17]. We find our optimal $T_A$, in terms of lowest $\rho_{xx}$ and highest T$_C$, to be ~ 250° C. As $T_A$ is further increased, we observe corresponding increase in $\rho_{xx}$ and decrease in $T_C$, as Mn leaves the Ga$_{1-x}$Mn$_x$As solution. For $T_A > $ ~350° C, $\rho_{xx}$ was too large to measure accurately, even at room temperature ($\rho_{xx} > $ 1 Ω-cm). For even the highest $T_A$, we have not detected any evidence of secondary precipitates such as $\tilde{\alpha}$-MnAs from SQUID magnetometry or HRXRD measurements. In short, by annealing three samples with nearly identical total Mn concentration, series of samples with resistivity spanning nearly two orders of magnitude, exhibiting both 'metallic' and insulator-like behaviors, and T$_C$ varying over 100 K are realized.

---

[a] Electronic mail: parkyd@phya.snu.ac.kr

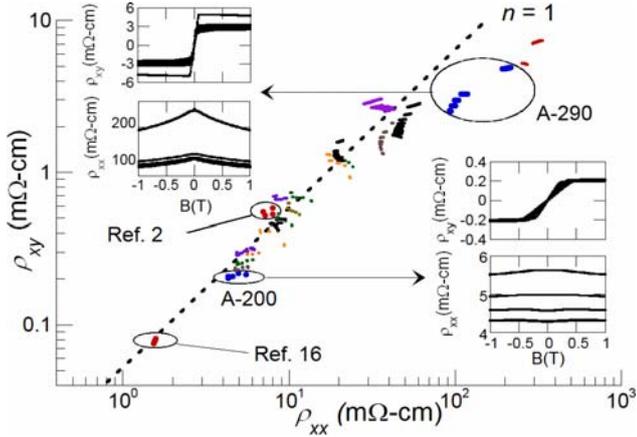
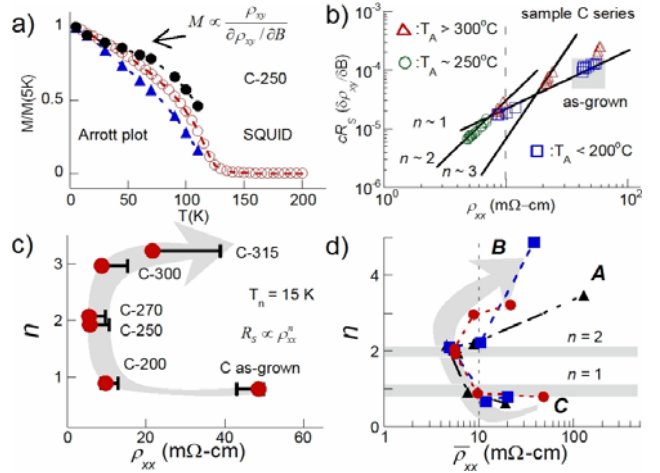

FIG. 1: Log-log plot of $\rho_{xy}$ and $\rho_{xx}$ for samples A-C series. Measurements of $\rho_{xy}$ and $\rho_{xx}$ are conducted simultaneously by ac lock-in technique with excitation current of 10 – 40 μA at 17 Hz. Each cluster of data points represents an isotherm of a particular sample from a series annealed at $T_A$ measured at $T_m$ ($< T_C$). Insets plot $\rho_{xy}$ and $\rho_{xx}$ as function of applied field for $T_m < T_C$ for sample A series annealed at 200° C (A-200) and at 290° C (A-290). Dotted line representing $n = 1$ is provides as a guide.

FIG. 2: a) Plot of sample C-250 normalized magnetization as function of temperature. b) Log-log plot of $cR_S$ ($\partial\rho_{xy}/\partial B$) as function of $\rho_{xx}$ for sample C annealed at 200° – 300° C to fit the scaling parameter $n$ ranging from ~1 to ~ 3. Each data point reflects isotherm measurements of AHE. c) Plot of scaling parameter $n$ as function of $\rho_{xx}$ at 15 K for sample C series (from as-grown to 315° C anneal). Arrow indicates increasing annealing temperature. 'Bars' span $\rho_{xx}$ measured up to $T_C$. d) Similar non-monotonic behavior of scaling parameter $n$ is seen in sample A (triangle) & B (square) series.

Hall measurements show typical anomalous behavior along with negative magneto-resistance for each sample below $T_C$. In AHE literature[18], the Hall resistivity ($\rho_{xy}$) is generally expressed as $\rho_{xy} = R_oB + \mu_oR_SM$, an empirical relationship valid for both intrinsic and extrinsic origins of magnetic material systems[19], where the first term is the ordinary term with $B$ as the magnetic induction and $R_o$ is the ordinary Hall coefficient related to the nature and amount of carriers, and the second term is the anomalous term with $M$ as the magnetization of the sample with $R_S$ related to spin polarization of carriers and the spin-orbit interaction. A scaling relationship for Hall resistivity can be expressed as $\rho_{xy} \propto \rho_{xx}^n$ in cases where magnetization is nearly constant. The value of the scaling parameter ($n$) can take on values of between one and two. Experimentally, $n$ can be determined by measuring $\rho_{xy}$ while varying $\rho_{xx}$ by *both* the measurement temperature ($T_m$) and solute concentration. Here, we vary the Mn concentration primarily by varying $T_A$. For all magneto-transport measurements, we plot $\rho_{xy}$ as function of $\rho_{xx}$ for each sample below its $T_C$ (Fig. 1). We note that our data along with those from other groups[2,16] generally follow a weakly universal linear scaling relationship ($n \approx 1$).

We apply an alternative scaling relationship in terms of the anomalous Hall coefficient ($R_S \propto \rho_{xx}^n$) as magnetic properties as well as $\rho_{xx}$ are inevitably related to carrier concentration ($n_h$). For Ga$_{1-x}$Mn$_x$As, the anomalous term is much larger than the ordinary term. Its influence along with difficulties in achieving technical magnetization saturation affects accurate determination of $n_h$ (along with accurate determination of $R_S$). We estimate $R_S$ at low applied fields from $\partial\rho_{xy}/\partial B=(\mu_oR_SM)\partial M_Z/\partial B_Z$ with $\mu_oR_SM \gg R_oB$ and $R_S$ being independent of $B$. For Ga$_{1-x}$Mn$_x$As during Hall measurement, $\partial M_Z/\partial B_Z$ can be expressed, from the Stoner-Wolfarth model, as $M_S[\mu_oM_S - 2(K_{u\perp} - K_C)]^{-1}$ here $K_{u\perp}$ and $K_C$ are the perpendicular uniaxial and the cubic anisotropy constants, respectively[20]. Recently, Titova *et al.* reports that the perpendicular uniaxial fields ($\mu_oM_S - 2 K_{u\perp}$) to be nearly independent of carrier concentration[21], and for $x > 0.05$, we expect $K_C$ to be negligible due to the large built-in compressive strain during LT-MBE[20]. A good agreement in the temperature dependence of normalized magnetization from SQUID magnetometer measurements, from Arrott plots of AHE data, as well as from $M \propto \rho_{xy}(\partial M_Z/\partial B_Z)^{-1}$ or $M \propto \rho_{xy}/R_S$ validates our assumption (Fig. 2.a).

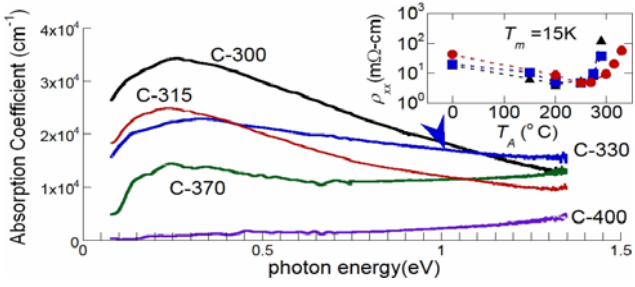

FIG. 3: Absorption spectra of sample series C annealed at differing temperatures (300° - 400° C) shows increased absorption near ~1 eV for samples annealed between 315° C and 370° C. Plot of $\rho_{xx}$ for sample series A (triangle), B (square), and C (circle) for differing annealing temperatures ($T_A$) (inset).

Now, we plot $cR_S$ (as $\partial\rho_{xy}/\partial B$) as function of $\rho_{xx}$, again for all samples and $T_m$ below sample's $T_C$ to fit $n$ (Fig. 2.b). We note a clear transition from $n = 2$ to $n = 1$ for $\rho_{xx} > \sim 10$ mΩ-cm, similar to what had been observed where only $\rho_{xx}$ had been varied by temperature[22]. For $\rho_{xx} < 10$ mΩ-cm, $n = 2$ is consistent with intrinsic origins of AHE[12,22,23]. For $n = 1$ regime, further study is required to discern whether the origins of AHE is due to hopping transport[13] or from extrinsic skew scattering[24] as linear scaling behavior is expected in both. For the highest $\rho_{xx}$, individually fitting the scaling relationship $R_S \propto (\rho_{xx}(T_m))^n$ where $\rho_{xx}$ is varied by measurement temperature results in atypical values of $n$ (> 3). We plot $n$ as fitted to each sample (Fig. 2.c&d). For all three series of samples, as-grown samples up to the optimal $T_A$ show $n \approx 1$. Near the optimal $T_A$, $n$ equals ~ 2, with further increase in $T_A$ results in $n > 3$, corresponding to decreases in $n_h$ and $T_C$ along with increase in $\rho_{xx}$. Similar scaling behavior of AHE has been seen in inhomogeneous granular systems where nanometer-sized super-paramagnetic clusters are randomly distributed in a non-magnetic matrix such as CoAg systems ($n = 3.7$)[25], although exact origins of such atypical scaling behavior are still controversial[26].

To determine whether our observations can be attributed to inclusion of metallic nanometer-sized particles, optical absorption measurements on sample C series are conducted (Fig. 3). Annealing the sample up to 300° C, the absorption increases and it is dominant near the low photon energy region. For $T_A > 300°$ C, the low photon energy absorption decreases which trend is consistent with dependence of $\rho_{xx}$ to $T_A$ from transport measurements (Fig. 3 inset). However, at $T_A$ of 330° C, it is noteworthy that the broad region of photon absorption around 1 eV increases. Such feature may originate from photon scattering by metallic clusters such as MnAs nano-crystals, whose diameters should be much smaller than the wavelength of the photon (1 eV = 1240 nm) and found to have average diameters of 20-30 nm from cross-sectional transmission electron microscopy in a previous study[27]. For annealing temperatures at 370° C and 400° C, overall photon absorption decreases. This observation suggests that the conducting carrier concentration dramatically decreases, and that the clusters size increases to a point that they cannot scatter photons by 'spherical resonance' process[28].

In summary, we have systematically measured the magneto-transport properties of annealed $Ga_{1-x}Mn_xAs$ ($x \approx 0.055$), which samples did not exhibit characteristics of secondary phases from SQUID magnetometry and HRXRD measurements. By determining the scaling relationships between the anomalous Hall coefficient and resistivity, samples annealed higher than 300° C exhibit scaling parameters that cannot be explained by current theories on the origins of AHE in DMS, and most likely due to formation of nanometer-sized metallic inclusions in a DMS matrix phase.

This work is supported by Samsung Electronics Endowment and KOSEF through CSCMR. YDP & KHK acknowledges partial support from the City of Seoul R&BD Program. We would like to thank H.C. Kim of MSL at KBSI for assistance with the SQUID magnetometry measurements.

[1] A. H. Macdonald, P. Schiffer, and N. Samarth, Nat. Mat. **4**, 195 (2005).
[2] H. Ohno, Science **281**, 951 (1998).
[3] M. Tanaka, J. Vac. Sci. Tech. B **16**, 2267 (1998).
[4] H. Ohno *et al.*, Nature (London) **408**, 944 (2000); Y. D. Park *et al.*, Science **295**, 651 (2002).
[5] Y.D. Park *et al.*, Phys. Rev. B **68**, 085210 (2003); A.M. Nazmul *et al.*, Phys. Rev. Lett. **95**, 017201 (2005).
[6] H. Toyosaki *et al.*, Nat. Mat. **3**, 221 (2004); S.R. Shinde *et al.*, Phys. Rev. Lett. **92**, 166601 (2004).
[7] A.P. Li *et al.*, Phys. Rev. B **72**, 195205 (2005).
[8] W. Allen *et al.*, Phys. Rev. B **70**, 125320 (2004).
[9] K. Hamaya *et al.*, Phys. Rev. Lett. **94**, 147203 (2005).
[10] V.F. Sapega *et al.*, Phys. Rev. Lett. **94**, 137401 (2005).
[11] K.Y. Wang *et al.*, Appl. Phys. Lett. **88**, 022510 (2006).
[12] T. Jungwirth, Q. Niu, and A. H. MacDonald, Phys. Rev. Lett. **88**, 207208 (2002).
[13] A.A. Burkov and L. Balents, Phys. Rev. Lett. **91**, 057202 (2003).
[14] Y.S. Kim *et al.*, J. Kor. Phys. Soc. **47**, 306 (2005).
[15] T. Hayashi *et al.*, Appl. Phys. Lett. **78**, 1691 (2001); S.J. Potashnik *et al.*, Appl. Phys. Lett. **79**, 1495 (2001).
[16] K.W. Edmonds *et al.*, Appl. Phys. Lett. **81**, 4991 (2002).
[17] K.W. Edmonds *et al.*, Phys. Rev. Lett. **92**, 037201 (2004).
[18] C.M. Hurd, *The Hall Effect and its applications*, edited by C. Chien and C.R. Westgate. (Penum, New York, 1980), pp.1-54.
[19] Z. Fang *et al.*, Science **302**, 92 (2003).
[20] X. Liu *et al.*, J. Appl. Phys. **98**, 063904 (2005).
[21] L.V. Titova *et al.*, Phys. Rev. B **72**, 165205 (2005).
[22] D. Ruzmetov *et al.*, Phys. Rev. B **69**, 155207 (2004).
[23] R. Karplus and J.M. Luttinger, Phys. Rev. **95**, 1154 (1954).
[24] J. Smit, Physica (Ultrecht) **21**, 877 (1955).
[25] P. Xiong *et al.*, Phys. Rev. Lett. **69**, 3220 (1992).
[26] A. Gerber *et al.*, Phys. Rev. B **69**, 224403 (2004).
[27] S.S.A. Seo *et al.*, J. Appl. Phys. **95**, 8172 (2004).
[28] S.S.A. Seo *et al.*, Appl. Phys. Lett. **82**, 4749 (2003).